\documentclass[twocolumn,twoside,dvips,showpacs,iop,epl,amsmath,amssymb]{revtex4}
\usepackage{bbm}
\usepackage{mathrsfs}
\usepackage{epsfig,psfrag}
\usepackage{amsmath,amsfonts,amssymb}
\usepackage[usenames]{color}
\usepackage{color}
\usepackage{soul}
\usepackage[dvips, bookmarks, colorlinks=true, plainpages = false, citecolor = blue, linkcolor = blue, urlcolor = blue, filecolor = blue]{hyperref}
\def\be{\begin{equation}}
\def\ee{\end{equation}}
\def\bea{\begin{eqnarray}}
\def\eea{\end{eqnarray}}

\begin{document}

\title{Growth models on the Bethe lattice}

\author{ Abbas Ali Saberi }\email{ab.saberi@ut.ac.ir}

\address {Department of Physics, University of Tehran, Post
  Office Box 14395-547, Tehran, Iran \\ Institut f\"ur Theoretische
  Physik, Universit\"at zu K\"oln, Z\"ulpicher Str. 77, 50937 K\"oln,
  Germany\\ Institute for Research in Fundamental Sciences (IPM),
School of Particles and Accelerators, Post Office Box 19395-5531,
Tehran, Iran}

\date{\today}

\begin{abstract}
I report on an extensive numerical investigation of various discrete
growth models describing equilibrium and nonequilibrium interfaces
on a substrate of a finite Bethe lattice. An unusual logarithmic
scaling behavior is observed for the nonequilibrium models
describing the scaling structure of the infinite dimensional limit
of the models in the Kardar-Parisi-Zhang (KPZ) class. This gives
rise to the classification of different growing processes on the
Bethe lattice in terms of logarithmic scaling exponents which depend
on both the model and the coordination number of the underlying
lattice. The equilibrium growth model also exhibits a logarithmic
temporal scaling but with an ordinary power law scaling behavior
with respect to the appropriately defined lattice size. The results
may imply that no finite upper critical dimension exists for the KPZ
equation.
\end{abstract}


\maketitle

The Kardar-Parisi-Zhang (KPZ) equation \cite{Kardar} is a simple
nonlinear Langevin equation that describes the macroscopic
properties of a wide variety of nonequilibrium growth processes
\cite{Stanley, Krug}. This equation is also related to many other
important physical problems such as the Burgers equation
\cite{Forster}, dissipative transport in the driven-diffusion
equation \cite{Beijeren} and directed polymers in a random medium
\cite{Huse, Imbrie, Parisi0}. The KPZ equation for a stochastically
growing interface described by a single valued height function
$h$(\textbf{x}, $t$) on a $d$-dimensional substrate \textbf{x}, is
\be\label{Eq1}\partial_t h(\textbf{x},t)=\nu \nabla^2
h+\frac{\lambda}{2}(\nabla h)^2+\eta(\textbf{x},t),\ee where the
first term represents relaxation of the interface caused by a
surface tension $\nu$, the second describes the nonlinear growth
locally normal to the surface, and the last is an uncorrelated
Gaussian white noise in both space and time with zero average $
\langle\eta(\textbf{x},t)\rangle=0 $ and $\langle
\eta(\textbf{x},t)\eta(\textbf{x}',t')\rangle=2D\delta^d(\textbf{x}-\textbf{x}')\delta(t-t')$,
mimicking the stochastic nature of the growth process. The steady
state interface profile is usually described in terms of the
roughness: $w=\sqrt{\langle h^2(\textbf{x},t)\rangle-\langle
h(\textbf{x},t)\rangle^2}$ which for a system of size $L$ behaves
like $L^\alpha f(t/L^{\alpha/\beta})$, where $f(x)\rightarrow$ const
as $x\rightarrow\infty$ and $f(x)\sim x^\beta$ as $x\rightarrow 0$,
so that $w$ grows with time like $t^\beta$ until it saturates to
$L^\alpha$ when $t\sim L^{\alpha/\beta}$. $\alpha$ and $\beta$ are
the roughness and the growth exponents, respectively, whose exact
values are known only for the special case $d=1$ as $\alpha=1/2$ and
$\beta=1/3$. The ratio $\bar{z}=\alpha/\beta$ is called dynamic
exponent. A scaling relation $\alpha+\bar{z}=2$ follows from the
invariance of Eq. (\ref{Eq1}) to an infinitesimal tilting of the
surface which retains only one independent exponent, say $\alpha$,
in the KPZ dynamics.

It is well known that for dimensions $d\leq2$ the surface is always
rough, while for $d>2$, the equation (\ref{Eq1}) shows two different
regimes in terms of the dimensionless strength of the nonlinearity
coefficient whose critical value $\lambda_c$ separates flat and
rough surface phases. In the weak coupling (flat) regime
($\lambda<\lambda_c$) the nonlinear term is irrelevant and the
behavior is governed by the $\lambda=0$ fixed point i.e., the linear
Edward-Wilkinson (EW) equation \cite{EW}, for which the exponents
are known exactly: $\alpha=(2-d)/2$ and $\beta=(2-d)/4$. In the more
challenging strong-coupling (rough) regime ($\lambda>\lambda_c$),
where the nonlinear term is relevant, the behavior of the KPZ
equation is quite controversial and characterized by anomalous
exponents. There is, however, a longstanding controversy concerning
the existence and the value of an upper critical dimension $d_c$
above which, regardless of the strength of the nonlinearity, the
surface remains flat.

At odds with many theoretical discussions supporting the existence
of a finite upper critical dimension \cite{Cook} between three and
four \cite{Bouchaud, Katzav}, and an analytical evidence that $d_c$
is bounded from above by four \cite{Lassig}, or many others
suggesting $d_c\approx2.5$ \cite{Doussal} or $d_c=4$ \cite{Colaiori,
Fogedby}, there is nevertheless a long list of evidence questioning
these suggestions \cite{Perlsman, Castellano, Ala-Nissila1,
Ala-Nissila2, Kim, Parisi, Schwartz1}, some of which concluded that
no finite upper critical dimension exists at all (for the most
recent study, see \cite{Schwartz1}).

Here I study the infinite dimensional properties of growth models
from two different KPZ and EW classes and compare them to realize
whether the nonlinear term in (\ref{Eq1}) is relevant in this limit.
It is inspired by the fact that if the nonlinear term is irrelevant
in infinite dimensions, then one would expect the same statistical
behavior for the models coming from each of the two classes. The
result would shed a light on the existence of the upper critical
dimension for the KPZ equation. To this aim, I investigate two
discrete nonequilibrium models, ballistic deposition (BD) and
restricted solid-on-solid (RSOS) models which are believed to be in
the KPZ class \cite{Stanley, Schwartz2, Kim2, Saberi1, Saberi2,
Saberi3}, as well as an equilibrium model, random deposition with
surface relaxation (RDSR) \cite{Family}, which belongs to the EW
class, all defined on the Bethe lattice, an effectively infinite
dimensional lattice. I find that the models from different
universality classes correspond to different statistical growth
properties and scaling behavior, the evidence that questions the
existence of a finite upper critical dimension for the KPZ equation.

\begin{figure}[t]
  \[
  \includegraphics[width=0.35\textwidth]{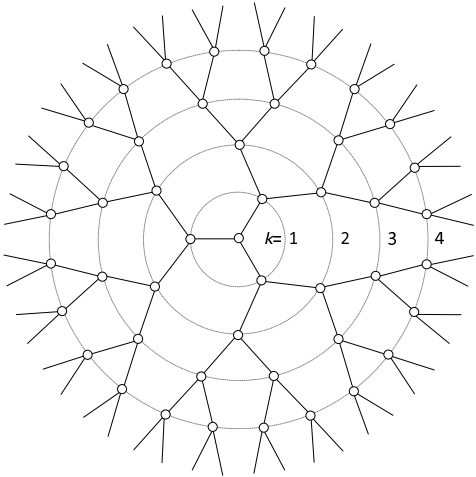}
  \]
  \caption{Part of a Bethe lattice with coordination
  number $z=3$ embedded in the plane which is considered here as a substrate of different growth
  models. The vertically incident particles can land at the top of the lattice sites
  represented by open small circles at different shells $k=0, 1, 2,
  \cdots$. For a given finite lattice of fixed size $k$, one lattice site is randomly
  chosen at each step and a particle is added to that site which can either increase the
  height according to the standard rules of BD and RSOS models, or it can diffuse through the neighboring
  edges until it finds the column with a
local minima in the searched area according to the RDSR model.
    \label{Fig0}}
\end{figure}

Due to its distinctive topological structure, several statistical
models involving interactions defined on the Bethe lattice
\cite{Bethe} are exactly solvable and computationally inexpensive
\cite{Baxter}. Various systems including magnetic models
\cite{Bethe}, percolation \cite{Thorpe, Stauffer, Chae, Sahimi0},
nonlinear conduction \cite{Sahimi1}, localization \cite{Thorpe,
Zirnbauer}, random aggregates \cite{Krug0, Bradley} and diffusion
processes \cite{Sahimi, Cassi, Alimohammadi} have been studied on
the Bethe lattice whose analytic results gave important physical
insights to subsequent developments of the corresponding research
fields. The Bethe lattice is defined as a graph of infinite points
each connected to $z$ neighbors (the coordination number) such that
no closed loops exist in the geometry (see Fig. \ref{Fig0}). A
finite type of the graph with boundary is also known as a Cayley
tree and possesses the features of both one and infinite dimensions:
since $N_k$, the total number of sites in a Bethe lattice with $k$
shells, is given as $N_k=[z(z-1)^k-2]/(z-2)$, the lattice dimension
defined by $d=\lim_{k\rightarrow\infty}[\ln N_k/\ln k]$ is infinite.
It is therefore often mentioned in the literature that the Bethe
lattice describes the infinite-dimensional limit of a hypercubic
lattice. As the lattice grows the number of sites in the surface, or
the last shell, grows exponentially $z(z-1)^{k-1}$. Therefore, as
the number of shells tends to infinity, the proportion of surface
sites tends to $(z-2)/(z-1)$. By surface boundary we mean the set of
sites of coordination number unity, the interior sites all have a
coordination number $z$. Thus the vertices of a Bethe lattice can be
grouped into shells as functions of the distances $k$ from the
central vertex. Here $k$ is the number of bonds of a path between
the shell and the central site and will be used as a measure of
lattice size.

\begin{figure}[b]
  \[
  \includegraphics[width=0.45\textwidth]{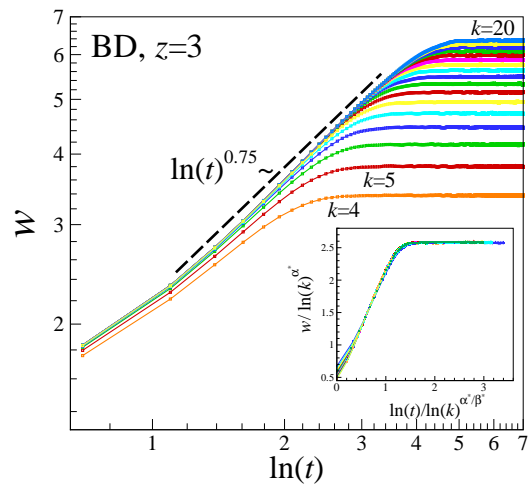}
  \]
  \caption{(Color online) Surface width $w$($t$,$k$) for the BD model
on a finite Bethe lattice of coordination number $z=3$, as a
function of logarithm of time, for the seventeen different sizes,
from the $k=4$th to the $20$th generation. Inset: Data collapse for
the same data with $k>6$. The time is rescaled by
$\ln(k)^{\bar{z}^\ast}$ (having assumed
$\bar{z}^\ast=\alpha^\ast/\beta^\ast=0.825/0.75=1.1$), and the width
is rescaled by $\ln(k)^{\alpha^\ast}$ with $\alpha^\ast=0.825$.
 \label{Fig1}}
\end{figure}

I have carried out extensive simulations of the BD, RSOS and RDSR
models on a finite Bethe lattice of different size $k$ and different
coordination number $z$ (Fig. \ref{Fig0}). I will first compute the
surface width $w$($t$,$k$) as a function of time $t$ and examine its
various scaling properties. For a given lattice size $k$, each Monte
Carlo time step is defined as the time required for $N_k$ particles
to deposit on the surface. I show that the surface widths for the
models feature a normal behavior as for a typical growth model on a
regular lattice: $w$ increases fast and finally saturates to a fixed
value $w_s$. Nevertheless, the best fit to our data at early time
before saturation shows that $w$ does not increase algebraically
with time ($w\sim t^{\beta}$), as is usually observed for growth
models on ordinary lattices. Rather I find a logarithmic scaling
behavior $w\sim \ln(t)^{\beta}$\footnote{For the sake of simplicity,
I use the same symbols for the exponents $\alpha$ and $\beta$, in
the logarithmic scaling laws.} for all considered models. I also
find that the saturated width $w_s$ of the interface for two BD and
RSOS models behaves like a logarithmic scaling law $w_s\sim
\ln(k)^{\alpha}$, while for the equilibrium RDSR model, it shows an
ordinary power law behavior with the lattice size, $w_s\sim
k^{\alpha}$. The model-dependent exponents are also found to be
functions of the coordination number of the underlying lattice,
$\alpha^{i}(z)$ and $\beta^{i}(z)$, where i= $\ast$, $\star$ and
$\circ$ denotes for BD, RSOS and RDSR models, respectively. Let me
call $\alpha^{i}(z)$ and $\beta^{i}(z)$, roughness and growth
exponents, respectively. This different scaling form with respect to
the finite-dimensional case can be associated to the exponential
(instead of polynomial) growth of the volume of a shell as a
function of its radius on the Bethe lattice.

I first consider the BD model on a lattice with $z=3$. Fig.
\ref{Fig1} shows the surface width $w$($t$,$k$) as a function of
logarithm of time, for the seventeen different sizes, from the $4$th
to the $20$th generation. At early times before saturation, the data
falls onto a straight line in a log-log scale indicating that the
surface width initially increases algebraically with the logarithm
of time as $w\sim \ln(t)^{\beta^\ast}$, with
$\beta^\ast(z=3)\simeq0.75(2)$. The best fit to the saturated width
$w_s$ as a function of different lattice size, gives a scaling
relation $w_s\sim \ln(k)^{\alpha^\ast}$, with
$\alpha^\ast(z=3)\simeq0.825(10)$.\\ As shown in the inset of Fig.
\ref{Fig1}, by standard rescaling of the parameteres, all curves
collapse onto a single function.

\begin{figure}[t]
  \[
   \includegraphics[width=0.4\textwidth]{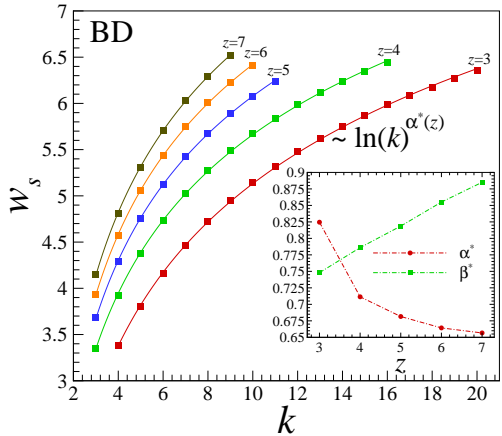}
   \]
   \caption{(Color online) Main: saturated surface widths as functions of lattice
   size $k$, for the BD model on the finite Bethe lattices of different
   coordination number $z=3$, $4$, $5$, $6$ and $7$. The solid lines
   show the best logarithmic fits of form $w_s\sim \ln(k)^{\alpha^\ast(z)}$ to the
   data. Inset: the growth $\alpha^\ast$ and roughness $\beta^\ast$
   exponents as functions of $z$.
    \label{Fig2}}
\end{figure}

In order to see how these exponents depend on $z$, series of
extensive simulations were performed for the BD model on a Bethe
lattice with different coordination number $z=3$, $4$, $5$, $6$ and
$7$. For each $z$, the surface width $w$($t$,$k$) was measured for
different lattice size $k$. Fig. \ref{Fig2} illustrates the
saturated surface widths as functions of the lattice size for each
coordination number. The solid lines in the figure show the best
logarithmic fits of form $w_s\sim \ln(k)^{\alpha^\ast(z)}$ to the
data, assigning a $z$-dependent roughness exponent $\alpha^\ast$ to
each data set. I also find that the growth exponent $\beta^\ast$ is
dependent on the coordination number of the substrate lattice.
$\alpha^\ast(z)$ and $\beta^\ast(z)$ are plotted in the inset of
Fig. \ref{Fig2}. As can be seen, $\alpha^\ast$ decreases, while
$\beta^\ast$ increases almost linearly with $z$.

\begin{figure}[b]
  \[
  \includegraphics[width=0.45\textwidth]{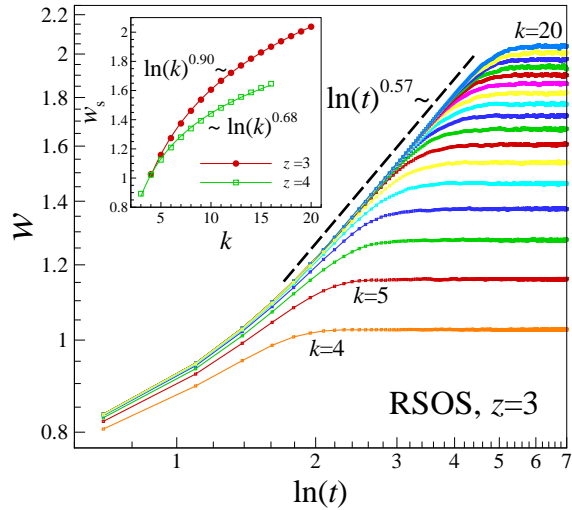}
  \]
  \caption{(Color online) Main: surface width $w$($t$,$k$) for the RSOS model
on a finite Bethe lattice of coordination number $z=3$, as a
function of logarithm of time, for the seventeen different sizes,
from the $k=4$th to the $20$th generation. Inset: saturated surface
widths as functions of lattice size $k$, for two different
   coordination numbers $z=3$ and $4$.
    \label{Fig3}}
\end{figure}

To see whether such a logarithmic scaling behavior is a
characteristic feature of the nonequilibrium growth models on the
Bethe lattice, I have also measured the surface width for the RSOS
model, for the seventeen different sizes, from the $k=4$th to the
$20$th generation, and for $z=3$. The results are shown in Fig.
\ref{Fig3}. These suggest the same scaling behavior but with
different estimated roughness and growth exponents
$\alpha^\star(z=3)\simeq0.90(1)$ and
$\beta^\star(z=3)\simeq0.57(2)$, respectively. The exponents for
this model depend again on the coordination number but both are
decreasing with $z$. For $z=4$, the exponents are estimated as
$\alpha^\star(z=4)\simeq0.68(1)$ and $\beta^\star(z=4)\simeq0.47(2)$
(see inset of Fig. \ref{Fig3}).

\begin{figure}[t]
  \[
  \includegraphics[width=0.5\textwidth]{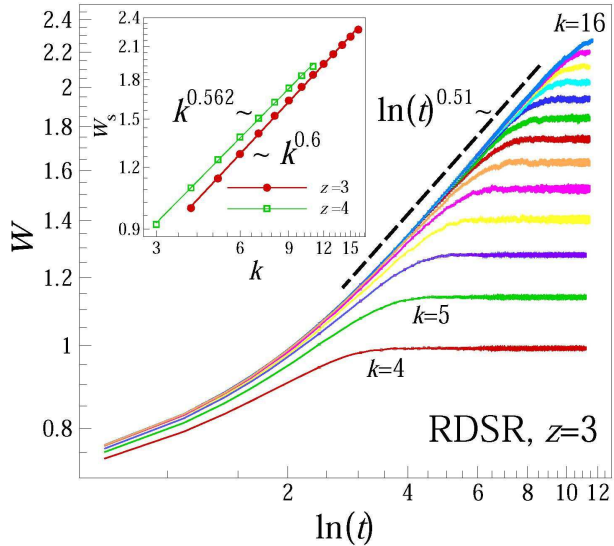}
  \]
  \caption{(Color online) Main: surface width $w$($t$,$k$) for the equilibrium RDSR model
on a finite Bethe lattice of coordination number $z=3$, as a
function of logarithm of time, for the thirteen different sizes,
from the $k=4$th to the $16$th generation. Inset: saturated surface
widths as functions of lattice size $k$, for two different
   coordination numbers $z=3$ and $4$. Unlike the nonequilibrium BD
   and RSOS models, $w_s$ here shows a power law scaling behavior $w_s\sim k^{\alpha^\circ(z)}$, with the lattice
   size $k$.
    \label{Fig4}}
\end{figure}

In random deposition with surface relaxation (RDSR) \cite{Family,
Meakin}, each particle is randomly dropped onto the surface, and it
is allowed to diffuse around on the surface within a prescribed
region about the deposited column, until it finds the column with a
local minima in the searched area. The corresponding continuum model
is the EW equation \cite{EW}.\\Fig. \ref{Fig4} summarizes the
results obtained from implementing the RDSR model on a finite Bethe
lattice of different size from the $k=4$th to the $16$th generation
with $z=3$. The first remarkable observation is that the crossover
time to the steady state is quite larger than that needed for the
above discussed nonequilibrium models, growing exponentially with
$k$ in this case. Therefore, the CPU time required for simulations
to reach the desired accuracy is orders of magnitude higher. As
shown in Fig. \ref{Fig4}, the temporal logarithmic scaling behavior
for this model again holds. I find the scaling relation $w\sim
\ln(t)^{\beta^\circ(z)}$, with $\beta^\circ(z=3)\simeq0.51(2)$.

The most remarkable scaling feature observed in the RDSR model is
that, unlike the two BD and RSOS models, the average saturated width
$w_s$ has no longer a power law relation with the logarithm of the
size, but with the size $k$ itself. The resulting data is plotted in
the inset of Fig. \ref{Fig4}. I find that $w_s\sim
k^{\alpha^\circ(z)}$, with $\alpha^\circ(z=3)\simeq0.60(1)$. The
roughness and growth exponents again depend on the coordination
number. Simulations for the RDSR model on a finite Bethe lattice of
different size from $k=3$th to the $14$th generation for $z=4$,
provide a satisfactory estimation of the exponents:
$\alpha^\circ(z=4)\simeq0.562(10)$ (see inset of Fig. \ref{Fig4})
and $\beta^\circ(z=4)\simeq0.46(2)$.

To summarize, I have studied three different growth models on a
substrate of a finite Bethe lattice with different coordination
number. A different scaling behavior is seen with respect to the
same models on the ordinary lattice or those on the fractal
substrates \cite{Lee}. Two models i.e., the BD and RSOS models, are
chosen from the nonequilibrium growth processes in the KPZ
universality class, and the third i.e., the RDSR model, is an
equilibrium model from EW class. For all considered models, the
surface width grows with a power law scaling relation with the
logarithm of the time before saturation. The initial growth is
characterized by an exponent which depends on the coordination
number of the underlying Bethe lattice as well as on the model in
question. In the steady state regime, the scaling behavior
distinguishes between equilibrium and nonequilibrium models. The
average saturated width for the nonequilibrium models has a power
law scaling relationship with the logarithm of the lattice size,
while for the equilibrium RDSR model, it shows a usual power law
behavior with the lattice size (instead of the logarithm of the
size).

If we admit that the Bethe lattice, as a substrate of a growth
model, reflects the infinite-dimensional limit properties of the
models, the present results would then imply that the nonlinear term
in the KPZ equation has a relevant contribution at this limit,
consequently questioning the existence of a finite upper critical
dimension for the KPZ equation.

I would like to thank J. Krug and M. Sahimi for their useful
comments, and H. Dashti-Naserabadi for his help with programming.
Supports from the Deutsche Forschungsgemeinschaft via SFB/TR 12, and
the Humboldt research fellowship are gratefully acknowledged. I also
acknowledge partial financial supports by the research council of
the University of Tehran and INSF, Iran.

\end{document}